\documentstyle[11pt,paspconf,epsf]{article}
\input psfig
\begin{document}

\title{Primordial Deuterium and Big Bang Nucleosynthesis:\\
      A Tale of Two Abundances}

\author{Gary\ Steigman}
\affil{Departments of Physics and Astronomy, The Ohio State University,
Columbus, OH 43210}


\begin{abstract}
Recent confrontations of the predictions of standard big bang 
nucleosynthesis (SBBN) with the primordial abundances of the 
light nuclides inferred from observational data reveal a conflict.  
Simply put, compared to theoretical expectations the inferred 
primordial abundances of either deuterium or helium-4 (or both) 
are ``too small".  Here I outline the ``tension" between D and 
$^4$He in the context of SBBN.  The incipient crisis for SBBN may 
be resolved by observations of deuterium in nearly pristine 
environments such as the high-redshift, low-metallicity QSO 
absorbers.  At present the big bang abundances of deuterium 
inferred from such data fall into two, apparently mutually 
exclusive, groups.  I describe the deuterium dichotomy and 
its implications for SBBN as well as for cosmology in general.

\end{abstract}




\section{Introduction}

The Universe is observed to be expanding and is filled with radiation 
(CBR at 2.7K) and ordinary matter (baryons).  In the past, when the 
density and temperature (thermal energy) were much higher, conditions 
may have permitted the synthesis of complex nuclei starting from 
neutrons and protons as the basic building blocks.  Which nuclei are 
synthesized and in what relative abundances depends on the details of 
the competition among the radiation density, the matter density and 
the early expansion rate.  In the context of the simplest, ``standard", 
hot big bang cosmology only the lightest nuclides (D, $^3$He, $^4$He 
and $^7$Li) are produced with astrophysically interesting yields.  
Furthermore, in SBBN these yields depend on only one ``free" parameter, 
the ratio of the nucleon (baryon) density to the photon density (today): 
$\eta \equiv n_{\rm B}/n_{\gamma}$ ; $\eta_{10} \equiv 10^{10}\eta$.  
In Figure 1 the predicted SBBN yields (for D, $^3$He and $^7$Li relative 
to hydrogen by number and for Y$_{\rm P}$, the $^4$He mass fraction) 
are shown as functions of $\eta$. 
\begin{figure}
\psfig{file=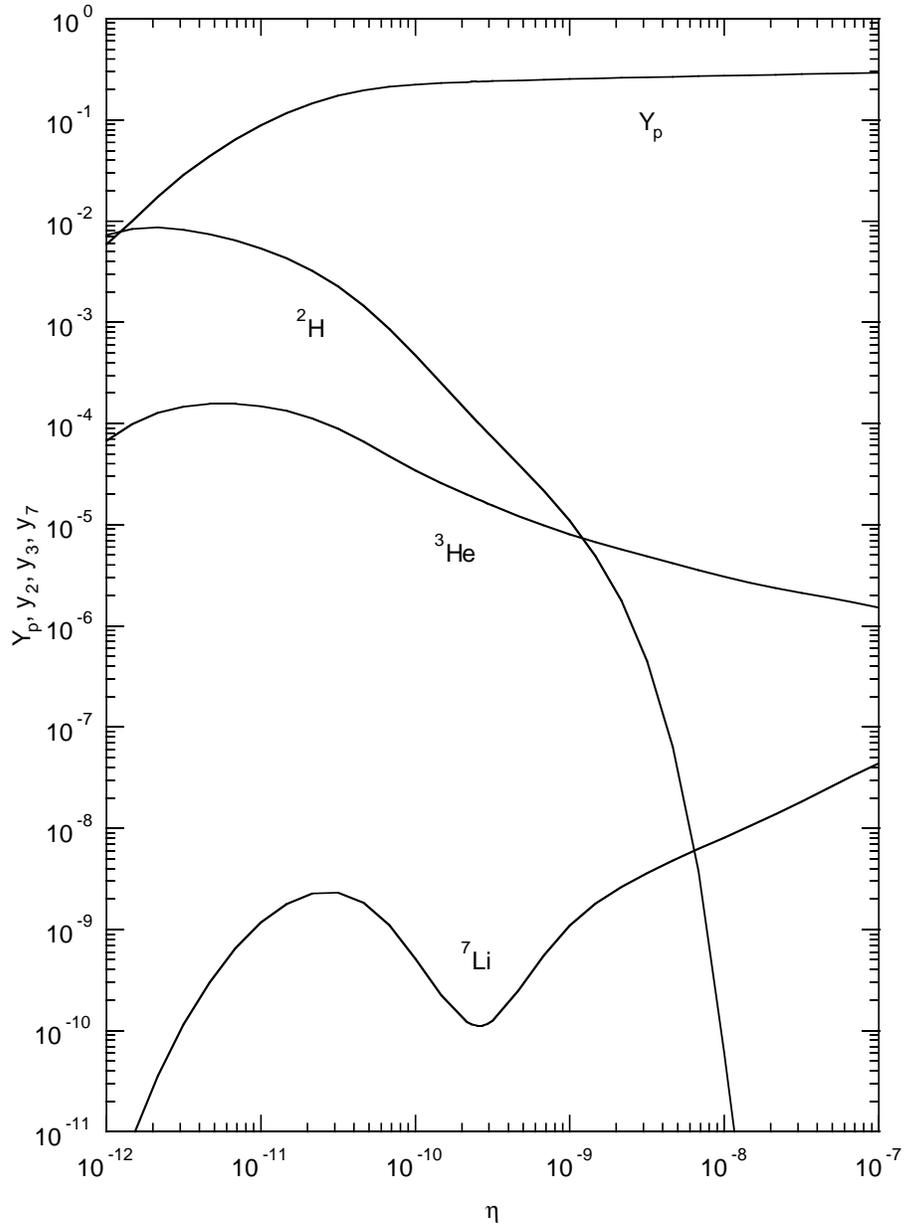,height=6.0in,width=4.5in}
\caption{SBBN-predicted abundances of $^4$He (mass fraction Y$_{\rm P}$), 
D ($y_{2\rm P}$ = (D/H)$_{\rm P}$), $^3$He ($y_{3\rm P}$ = 
($^3$He/H)$_{\rm P}$) and $^7$Li ($y_{7\rm P}$ = ($^7$Li/H)$_{\rm P}$) 
as functions of $\eta$, the present ratio of nucleons to photons.  
This graph has been provided by D. Thomas.} 
\label{fig-1}
\end{figure}
Clearly SBBN is - in principle - an overdetermined theory, 
predicting four primordial abundances at the expense of one 
parameter.  If the primordial abundance could be determined 
for only one of the light nuclides $\eta$ would be fixed and 
SBBN would predict the abundances of the other three.  Confrontation 
between such predictions and the observational data constitutes 
a test of the standard model.  In practice this approach has 
encountered many obstacles.

For example, until recently deuterium has been observed ``here and now" 
in the local interstellar medium (ISM) or in the solar system.  The 
good news is that the evolution of D from the big bang (``there and then") 
is straightforward: 
In the course of Galactic evolution D is destroyed (burned to $^3$He and 
beyond) whenever gas is incorporated into stars.  Thus the abundance 
(mass fraction X$_{2}$) of deuterium has only decreased since the big 
bang (X$_{2\rm P} \geq$ X$_{2}$).  The bad news is that, in the absence 
of reliable constraints on the amount of chemical evolution (up to the 
present or to the time the solar nebula formed), we don't know by how 
much.  Conventional models of chemical evolution, constrained by the size 
and distribution (in space as well as in time) of various heavy-element 
abundances, predict only modest D-destruction by a factor of 
2 -- 3 or so (e.g., Audouze \& Tinsley 1974; Vangioni-Flam \& Audouze 
1988; Tosi 1988a,b, 1996; Steigman \& Tosi 1992, 1995; Fields 1996).  
If so, current data point towards a ``low" primordial abundance implying 
a ``high" $\eta$ (see Fig. 1).  However, unconventional models 
have been designed (e.g., Scully et al.\ 1996) which arrange for 
much more D-destruction while maintaining consistency with the 
other observational data.  If this, or something like it, is the 
path chosen by the Galaxy the primordial abundance of D might be 
much larger than current data suggest, permitting a smaller 
nucleon abundance (see Fig. 1). 

The extraction of the primordial abundance of $^3$He from solar-system 
and/or ISM (H II regions) data (Bania, Rood \& Wilson 1987; 
Rood, Bania \& Wilson 1992; Balser et al.\ 1994) is even more 
problematic (e.g., Dearborn, Steigman \& Tosi 1996) given that 
some $^3$He is destroyed, some survives, and new $^3$He is
produced in the course of the evolution of stars of different 
masses.  The delicate balance between stellar and Galactic 
evolution renders challenging in the extreme an accurate 
inference of the primordial abundance.  However, some
qualitative features have emerged from such attempts.  By 
the time stars reach the main sequence they have already 
converted any prestellar deuterium to helium-3; the more D 
destroyed, the more $^3$He has passed through stars.  Since 
not all the pre-main sequence helium-3 (D + $^3$He) is 
destroyed (by stars of any mass), there is some correlation 
between the observed abundances of D and $^3$He and their 
primordial values (e.g., Yang et al.\ 1984; Walker et al.\ 
1991; Steigman \& Tosi 1992, 1995).  Consequently, in most 
analyses of SBBN it is not $^3$He but some combination of D 
and $^3$He which provides a means of testing for consistency.

Although not without problems of its own, in some ways lithium is 
simpler.  Since the pioneering work of the Spites (Spite \& Spite 
1982a,b,; Spite, Maillard \& Spite 1984; Spite \& Spite 1986) a 
large body of observations of lithium in very old, very metal-poor 
Pop II halo stars has been accumulated (Rebolo, Molaro \& Beckman 
1988; Hobbs \& Pilachowski 1988; Thorburn 1994; Molaro, Primas \& 
Bonifacio 1995; Spite et al.\ 1996).  The very low metallicity 
guarantees that the extrapolation to primordial is minimal.  But 
the very great ages of these stars open the possibility that 
they may have altered their surface abundance of $^7$Li from 
its prestellar value (e.g., Chaboyer et al.\ 1992; Pinsonneault, 
Deliyannis \& Demarque 1992; Charbonnel \& Vauclair 1992, 1995).  
Further, since the large number of careful, independent 
observations ensures high statistical accuracy, it is uncertainties 
in the model atmospheres and the temperature scale for these 
metal-poor stars which dominate the abundance errors.  
Nevertheless, few would disagree that the data point towards a low 
primordial abundance, considerably below the Pop I value (see, e.g., 
Steigman 1996), near the minimum of the ``lithium valley" (see Fig. 1).  
At present, uncertainties in the precise value of the ``Spite plateau" 
and in the amount of possible stellar dilution/destruction relegate 
$^7$Li to the role of offering an approximate rather than a precise 
test of SBBN (see, however, Fields \& Olive 1996; Fields et al.\ 1996): 
The inferred primordial lithium abundance is consistent with a range 
in nucleon abundance from ``low" to ``high" values of $\eta$ (see Fig. 1).

Helium-4 provides the last, best hope for testing SBBN quantitatively.  
As the second most abundant element (after hydrogen), $^4$He may be 
observed throughout the Universe (not just ``here and now").  And 
large numbers of careful observations may be used to determine its 
abundance to high statistical accuracy.  High-quality data for 
individual observations achieve 5\% accuracy ($\approx$ 0.012 in Y) 
and there now exist several dozen such determinations.  However, the 
flip side of the $^4$He coin is that the predicted primordial 
abundance (mass fraction, Y$_{\rm P}$) is a very weak function 
of the one free SBBN parameter ($\eta$) as is evident from Figure 1.  
Thus, to provide a key test of the consistency of SBBN, it is 
necessary to pin down Y$_{\rm P}$ very accurately.  Since stars 
do produce new $^4$He along with the heavy elements (``metals"), 
the correction for evolution must be minimized or understood very 
well.  For this reason most effort in the search for the holy 
grail (Y$_{\rm P}$) has been channelled towards observations of 
the low-metallicity, extragalactic H II regions (Searle \& Sargent 
1971; Peimbert \& Torres-Peimbert 1974; Lequeux et al.\ 1979; 
Kunth \& Sargent 1983; Torres-Peimbert, Peimbert \& Fierro 1989;
 Pagel et al. 1992; Skillman \& Kennicutt 1993; Skillman et al.\ 
1994; Izotov, Thuan \& Lipovetsky 1994).  It is in the 
interpretation of this body of data that
unknown (or unquantified) systematic errors may rear their ugly 
heads (Davidson \& Kinman 1985; Pagel et al.\ 1992; Sasselov \& 
Goldwirth 1995; Peimbert 1996; Skillman 1996).  Current data 
(see Olive \& Steigman 1995, 1996 and references therein) point 
towards a ``low" value of Y$_{\rm P}$, indicating a ``low" value 
of $\eta$ (see Fig. 1).  But an unaccounted-for systematic offset 
in the primordial mass fraction, $\Delta$Y$_{\rm sys}$, might raise 
Y$_{\rm P}$, and correspondingly, the upper bound to $\eta$.

\section{The Tension Between Deuterium and Helium-4}

Until recently the confrontation of the SBBN predictions with the 
primordial abundances inferred from increasingly extensive and 
accurate observational data has been a spectacular success, 
confirming the consistency of SBBN, zeroing in on the 
universal abundance of nucleons ($\eta$) and constraining 
non-standard models of particle physics (see, e.g., Walker 
et al.\ 1991 and references therein).  However, it has begun 
to be increasingly clear that the very tight constraints 
inferred from this confrontation of theory with observations 
are a precursor to the conclusion that the ``standard" model 
is not providing a very good fit to the data.  One way to 
describe this emerging crisis for SBBN is through the parameter 
N$_{\nu}$, the ``effective number of equivalent light neutrino 
flavors" (Steigman, Schramm \& Gunn 1977).  The early expansion 
rate of the Universe is controlled by the total mass-energy 
density which, at very early epochs, is dominated by the 
contribution from relativistic particles.  For SBBN these 
are photons ($\gamma$), electron-positron pairs (e$^{\pm}$), 
and provided they are light (i.e., relativistic), neutrinos 
($\nu_{e}$, $\nu_{\mu}$, $\nu_{\tau}$).  For SBBN the total 
density, in units of the photon density, is 
$\rho_{\rm TOT}^{\rm SBBN}/\rho_{\gamma}$ = 43/8.  A convenient 
way to parameterize any deviation of the early expansion rate 
from its SBBN value is through N$_{\nu}$, where

\begin{equation}
\rho_{\rm TOT}/\rho_{\rm TOT}^{\rm SBBN} \equiv 1 + 7(N_{\nu} - 3)/43.
\end{equation}
For SBBN, N$_{\nu}$ = 3 while if N$_{\nu}$ $>$ 3 the early universe 
expands more rapidly, leaving behind more neutrons to be incorporated 
into $^4$He during primordial nucleosynthesis; for N$_{\nu} 
\approx$ 3, $\Delta$Y$_{\rm P}$ $\approx$ 0.01(N$_{\nu} - 3$).  
A lower bound to $\eta$, for example from D and/or D+$^3$He, 
combined with an upper bound to Y$_{\rm P}$ provides an upper 
bound to N$_{\nu}$ of importance in constraining attempts to 
go beyond the standard model of particle physics (Steigman, 
Schramm \& Gunn 1977; Yang et al.\ 1979; Steigman, Olive \& 
Schramm 1979; Yang et al.\ 1984; Walker et al.\ 1991).  The 
closer this upper bound is to the SBBN value of 3, the more 
restrictive the constraint.  For example, Walker et al. (1991) 
found a ``95\% CL" constraint of N$_{\nu}$ $<$ 3.3, ``forbidding" 
a fourth generation of light neutrinos or even a new light scalar
 particle.  By 1994, Kernan \& Krauss had reduced this bound to 
3.04 which, if it is as claimed a ``95\% CL" upper bound, strongly 
suggests that SBBN (with N$_{\nu}$ = 3) is not a good fit to the 
data.  Indeed, Olive \& Steigman (1995) in their reanalysis of the 
H II region $^4$He data found a central value for N$_{\nu}$ of 2.2 
and a 2$\sigma$ (statistical) upper bound of 2.7, presaging the 
current crisis (Hata et al.\ 1995).  Only when they included a
 possible systematic offset of $\Delta$Y$_{\rm sys}$ = 0.005 did 
Olive \& Steigman (1995) recover consistency (barely) with SBBN 
(N$_{\nu}$ $<$ 3.1).  Of course the problem may not (only) be 
with Y$_{\rm P}$.  Perhaps the upper bound on D (or D+$^3$He) 
has been too severe, leading to a lower bound to $\eta$ (and 
therefore to the SBBN predicted value of Y$_{\rm P}$) which is 
too restrictive.

At the same time that the upper bound to Y$_{\rm P}$ was shrinking, 
tightening the noose around SBBN, the lower bound to $\eta$ was in 
fact becoming more -- not less -- restrictive (Steigman \& Tosi 1992, 
1995; Dearborn, Steigman \& Tosi 1996).  For a ``low" value of 
$\eta$ the SBBN-predicted primordial abundance of
D is ``high", corresponding to an SBBN-predicted primordial 
abundance for $^4$He which is ``low" (just what we ``want"; 
see Fig. 1).  But the observed ISM deuterium abundance 
(McCullough 1992; Linsky et al. 1993), as well as that 
inferred for the presolar nebula from solar system data 
(Geiss 1993), is ``low".
A high primordial abundance can only be reconciled with a 
low value ``here and now" if most of the gas in the present 
ISM (or in the presolar nebula) has been cycled through stars 
where D is destroyed.  But there are observational consequences 
of an efficient cycling of gas through stars.  Stars produce 
heavy elements and the metal abundance is observed to be small.  
Long-lived low-mass stars tie up gas, lowering the ratio of the 
mass in gas to the mass in stars which is not observed to be 
very small (Gould, Bahcall \& Flynn 1996).  Given these and 
similar restrictions on Galactic evolution most models of 
chemical evolution predict only modest D-destruction (e.g., 
Tosi 1996; Fields 1996 and references therein).  A somewhat 
more direct constraint on the magnitude of possible D-destruction 
may be inferred from the fact that when D is destroyed it is 
burned to $^3$He which is not entirely destroyed in stars 
(some of which -- the low mass ones -- are expected to be net
 producers of $^3$He).  Again, the results of conventional 
chemical evolution models (e.g., Steigman \& Tosi 1992; 
Dearborn, Steigman \& Tosi 1996; Fields 1996) as well as 
of model-independent ``inventories" (Steigman \& Tosi 1995; 
Hata et al. 1996a) suggest little D-destruction and a 
relatively low primordial abundance corresponding to 
relatively high values of $\eta$ and of Y$_{\rm P}$.  

\subsection{The Crisis}

This ``tension" between deuterium (favoring relatively high values 
of $\eta$ and Y$_{\rm P}$) and helium-4 (favoring a relatively low 
value of $\eta$ and a relatively high value of X$_{2\rm P}$) 
constitutes the ``crisis" for SBBN (Hata et al. 1995).  In Figure 
2 (from Hata et al. 1995) the inferred primordial abundances for 
D, $^4$He and $^7$Li are superposed on the SBBN predictions with 
each comparison delimiting its corresponding range of $\eta$.  
\begin{figure}
\psfig{file=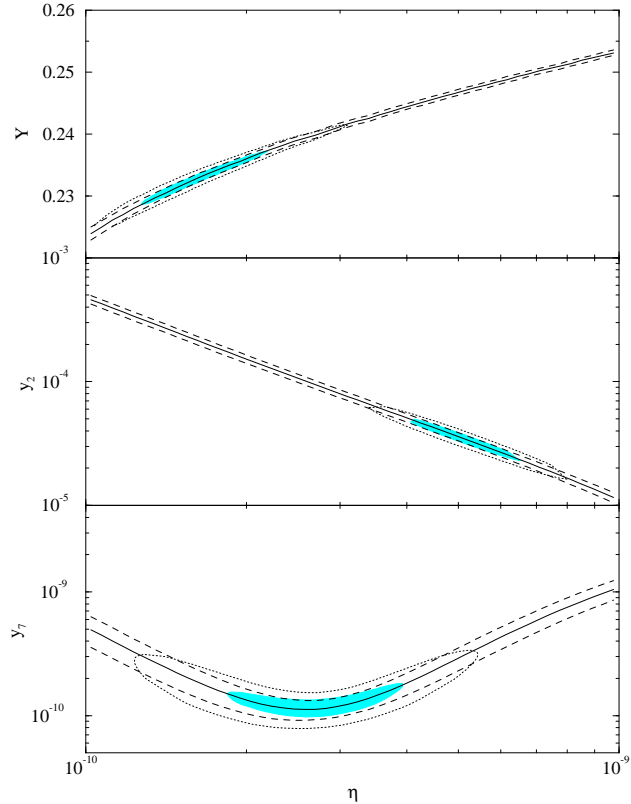,height=4.3in,width=3.23in}
\caption{SBBN predictions (solid lines) for Y$_{\rm P}$, $y_{2\rm P}$ 
and $y_{7\rm P}$ along with their 1$\sigma$ uncertainties estimated 
via Monte Carlos (dashed lines).  The regions constrained by 
observations at 68\% CL (shaded) and at 95\% CL (dotted lines) are 
also shown.  (From Hata et al. 1995.)} 
\label{fig-2}
\end{figure}
The challenge to SBBN posed by the data is that the ranges in $\eta$ 
determined from D do not overlap those from $^4$He (at ``95\% CL"); 
note that $^7$Li is consistent with either D or $^4$He while 
slightly preferring low $\eta$.  Another way to view the problem 
for SBBN is to use the primordial abundances inferred from the 
observational data to construct the likelihood distribution for 
N$_{\nu}$.  The best fit is for N$_{\nu} = 2.1 \pm 0.3$ (N$_{\nu}$ 
$<$ 2.6 at ``95\% CL").  A measure of the ``goodness of fit" of SBBN 
is provided by the ratio of likelihood at N$_{\nu}$ = 3 to that 
at 2.1, 0.014 (Hata et al. 1995); SBBN is not a good fit to the data.

\subsection{Paths To Resolution}

The crisis described above may be ``real" (SBBN might need to be 
modified) or ``illusory" (one or more of the primordial abundances 
inferred from the observational data might be in error).  If ``real", 
the simplest of many possibilities is that the tau neutrino 
is massive (Kawasaki et al. 1994).
For example, if m($\nu_{\tau}$) $\approx 10 - 20$ MeV, the tau 
neutrinos would not be fully relativistic at the epoch of BBN. 
This would modify the expansion rate, changing the number of 
neutrons available to be incorporated into $^4$He.  Such a massive 
neutrino species would have to be unstable (to avoid ``overclosing" 
the Universe soon after BBN) so that the 
deviations from the predictions of SBBN would depend on its 
lifetime and its decay products.  Kawasaki et al. (1994) 
considered the simple decay channel  $\nu_{\tau} \rightarrow 
\nu_{\mu} + \phi$, where $\phi$ is a light,``sterile", scalar particle.  
In this case, for 10 $\leq$ m($\nu_{\tau}$) $\leq$ 24 MeV and a 
lifetime from $\approx$ 0.01 sec. to $\approx$ 1 sec., an 
appropriate reduction in Y$_{\rm P}$ may be achieved.  Perhaps 
cosmology is pointing the way to new physics beyond the standard 
model of elementary particles.

If, instead, the crisis is ``illusory", many options present themselves 
for consideration.  Since the ``tension" is between D and $^4$He, it is 
natural to concentrate on each of them.  Perhaps, for example, the 
extragalactic H II
region data have led to an underestimate of the primordial helium 
abundance.
If, rather than the inferred value of Y$_{\rm P} = 0.232 \pm 0.003$ 
(Olive \& Steigman 1995), the true value were larger by $\Delta$Y 
$\approx$ 0.014, the crisis would be resolved.  The small statistical 
uncertainty in Y$_{\rm P}$, traceable to the large numbers of 
carefully observed H II regions, suggests that
only an unacknowledged systematic offset in the derived $^4$He 
abundances could
account for such a large difference.  Many of the potential 
sources of systematic error in Y would be expected to vary from 
source to source (e.g., ionization corrections, collisional 
excitation, stellar absorption, etc.) or from observer 
(telescope/detector) to observer (e.g., signal/noise, 
calibration, etc.) leading to a dispersion about the mean Y 
versus metallicity relation accompanying any systematic offset.  
The low reduced $\chi^{2}$ ($\la 1$) for the data compared to
a simple two-parameter fit (Olive \& Steigman 1995) suggests 
that any such offsets should be small unless all H II 
regions are shifted uniformly (e.g., in the unlikely case of 
incorrect atomic data).  Thus, although it would take 
only a 6\% upward shift in Y$_{\rm P}$ to resolve the SBBN 
crisis, it is difficult to identify any obvious source of 
such an error.  Perhaps, then, deuterium is the culprit.

Certainly any attempt to use ISM and/or solar system data on D 
(and $^3$He) to
infer the primordial yield is placed in peril by the necessary 
intervention of models of Galactic evolution.  Although conventional 
models suggest only modest
D-destruction (e.g., Tosi 1996; Fields 1996), ``designer" models
may permit much
more astration (Scully et al. 1996). If, rather than the typical 
factor of 2 -- 3 destruction, the ISM abundance (X$_{2\rm ISM}$) 
lies below the primordial mass fraction by a factor of 5 -- 10, 
the tension between D and $^4$He would be relieved.  Whether or not 
this is the case is better left to observations than to theory.  
If deuterium could be observed in very old, nearly pristine 
astrophysical environments, the primordial abundance could be 
determined without recourse to models of chemical evolution.  
The very high redshift ($z \approx$ $2 - 4$), very low metallicity 
($Z \leq Z_{\odot}$/100) QSO absorption-line systems provide ideal 
targets and data from observations of such systems are now becoming 
available (Songaila et al. 1994; Carswell et al. 1994; Rugers \& Hogan 
1996; Tytler, Fan \& Burles 1996; Burles \& Tytler 1996).  The remainder 
of this paper is devoted to exploring the implications for BBN (and 
for cosmology in general) of these new data.

\section{A Tale of Two Deuterium Abundances}

It is the best of times (a growing body of data on nearly primordial D 
from observations of high-$z$, low-$Z$ QSO absorbers); it is the worst 
of times (the
data appear inconsistent, leading to a deuterium dichotomy).  On the 
one hand, the data of Songaila et al. (1994), Carswell et al. (1994) 
and Rugers \& Hogan (1996) from two separate absorbing systems towards 
the same QSO yield a very high D abundance: D/H = 19 $\pm$ 4 $\times$ 
$10^{-5}$, while those of Tytler, Fan \& Burles (1996) and of Burles 
\& Tytler (1996) of two systems towards two different QSOs imply a 
much lower value: D/H = 2.4 $\pm$ 0.3 $\pm$ 0.3 $\times$ $10^{-5}$.  
Although other data are available at present, they only offer bounds 
to the deuterium abundance and I will not consider them further here.  
In Figure 3, very similar to Figure 2, are shown the two 
deuterium determinations in comparison to the predictions of 
SBBN; the $^4$He 
and $^7$Li plots are exactly as in Fig. 2.  
\begin{figure}
\psfig{file=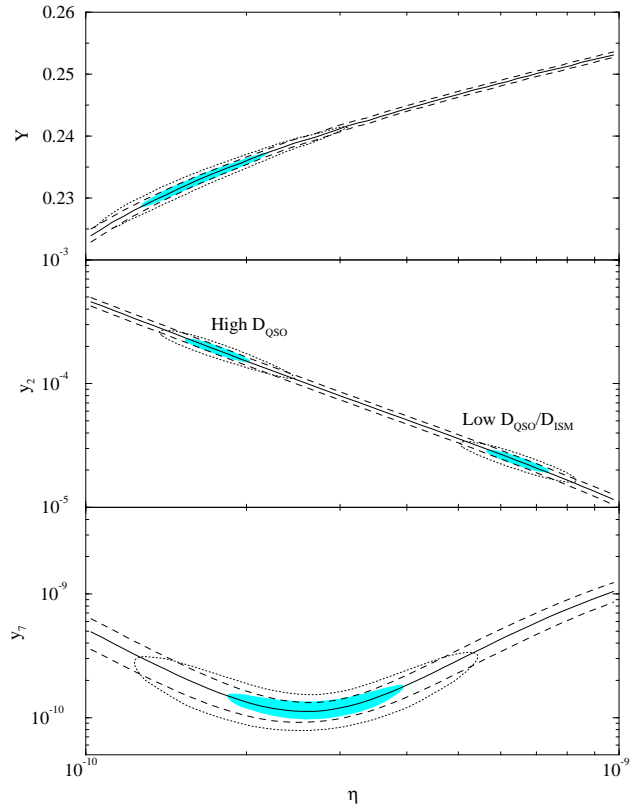,height=4.3in,width=3.23in}
\caption{As for Figure 2, but with the deuterium abundances inferred 
from the two sets of QSO absorption data.  (From Hata et al. 1996b.)} 
\label{fig-3}
\end{figure}
This figure permits us to explore the consequences for SBBN of the 
``high-D" and the ``low-D" observations.

\subsection{High-D}

Figure 3 shows clearly that if the ``high-D" abundance is primordial 
the tension between D and $^4$He is relieved and consistency for SBBN 
is recovered.  As Hata et al. (1996b) have found, this deuterium 
abundance selects (for SBBN) a ``95\% CL" range for $\eta$ of 1.3 
$\leq \eta_{10} \leq$ 2.7 corresponding to SBBN-predicted primordial 
abundances for $^4$He (0.231 $\leq$ Y$_{\rm P}$ $\leq$ 0.239) and for
 $^7$Li (0.7 $\leq 10^{10}$($^7$Li/H) $\leq$ 3.0) which are in
 excellent agreement with the primordial values inferred from the 
H II region and halo-star data respectively.  Alternatively, a 
likelihood analysis (Hata et al. 1996b) finds for the ``best value" 
of N$_{\nu}$, 2.9 $\pm$ 0.3, entirely consistent with SBBN 
(N$_{\nu} = 3$).  Corresponding to an upper bound to Y$_{\rm P}$ 
of 0.243 (which includes an allowance for a possible systematic 
offset in Y$_{\rm P}$ of 0.005; Olive \& Steigman 1995), ``high-D" 
sets an upper bound to N$_{\nu}$ of 3.6, forbidding the existence 
of a fourth generation of light neutrinos (or, its equivalent) but
permitting (just barely) a light scalar.

There are two problems associated with such a high value for the 
primordial abundance of deuterium.  Although perfectly consistent 
with SBBN, this high initial D abundance requires that 90\% or 
more of the present ISM has been cycled through stars (i.e., 
X$_{2\rm P}$/X$_{2\rm ISM} = 13 \pm 3$).  Can such efficient 
stellar processing occur without overproducing the present 
metallicity and/or using up the interstellar gas?  Conventional 
evolution models suggest no.  It will be a challenge to 
``designer" models to see if this is possible (consistent 
with observations).  Another challenge posed by the 
identification of ``high-D" with the primordial abundance 
of deuterium is the correspondingly low baryon density it 
implies.  In terms of the density parameter $\Omega_{\rm B}$ 
(the ratio of the present mass density in baryons to the 
``critical" density) and the normalized Hubble parameter 
$h$ ($H_0 = 100h$ km/sec/Mpc),

\begin{equation}
\Omega_{\rm B} h^2 = 3.66 \times 10^{-3}\eta_{10},
\end{equation}
so that for 1.3 $\leq \eta_{10} \leq$ 2.7, 0.005 $\leq \Omega_{\rm B} 
h^2 \leq$ 0.010.  This corresponds to a very low baryon density; for 
$H_0$ between 50 and 100 km/sec/Mpc, 0.005 $\leq \Omega_{\rm B} \leq$ 
0.040.  As we shall see shortly, this low baryon density may be in 
conflict with observations of the x ray-emitting hot gas in rich 
clusters of galaxies (White et al. 1993; Steigman \& Felten 1995; 
Hata et al. 1996b; Steigman, Felten \& Hata 1996).  Before turning to 
considerations of mass density, let us first consider the implications 
for SBBN of the ``low-D" abundance.

\subsection{Low-D}

If, indeed, the ``low-D" determinations reflect the primordial D 
abundance, SBBN is in trouble.  In this case the situation is 
somewhat worse than before (the tension between D and $^4$He is 
increased) due to the rather low value for X$_{2\rm P}$ implied 
by the ``low-D" QSO data.  This abundance, which leaves very 
little room for any D-destruction (X$_{2\rm P}$/X$_{2\odot} = 
1.0 \pm 0.4$; X$_{2\rm P}$/X$_{2\rm ISM} = 1.6 \pm 0.4$), 
identifies a high range for $\eta$ (5.1 $\leq \eta_{10} \leq$ 
8.2)  corresponding to high predicted (for SBBN) primordial 
abundances 
of $^4$He (0.246 $\leq$ Y$_{\rm P} \leq$ 0.252) and of $^7$Li 
(3.0 $\leq 10^{10}$($^7$Li/H) $\leq$ 7.8).  Now consistency 
with SBBN requires not only that the H II region observations have 
systematically underestimated Y by an amount of order 0.017 (7\%), 
but also that the ``Spite plateau" halo stars should
have destroyed or diluted their initial lithium by more than a 
factor of three.  In contrast to these apparently serious problems, 
the baryon density corresponding to ``low-D" is quite reasonable, 
0.019 $\leq \Omega_{\rm B} h^2 \leq$ 0.030 (for $1/2 \leq h \leq 1$, 
0.12 $\geq \Omega_{\rm B} \geq$ 0.019).   

\subsection{N$_{\nu}$ Revisited}

If the QSO D-abundances are used to constrain $\eta$ and N$_{\nu}$ 
is allowed to
be a free parameter, the BBN-predicted abundance of $^4$He will 
depend on N$_{\nu}$.  We have seen that for ``high-D" and for 
$\Delta$Y$_{\rm sys}$ = 0, SBBN (N$_{\nu}$ = 3) is a good fit.  
However, as $\Delta$Y$_{\rm sys}$ is increased with $\eta$ fixed 
by ``high-D", the ``best-fit" value of N$_{\nu}$ also increases.  
Eventually, for sufficiently large $\Delta$Y$_{\rm sys}$, 
N$_{\nu}$ = 3 is no longer a good fit and SBBN is in trouble.  
This behavior is shown in Figure 4 from Hata et al. (1996b). 
\begin{figure}
\psfig{file=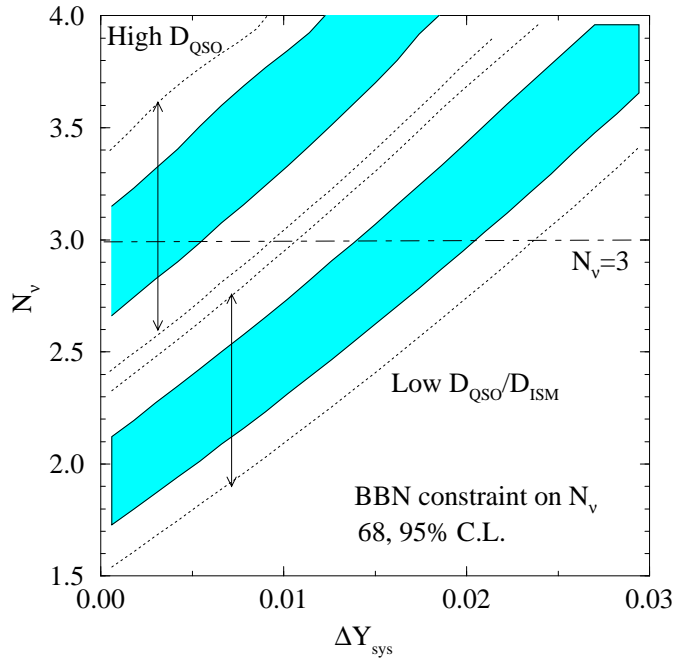,height=3.8in,width=4.1in}
\caption{The allowed range of N$_{\nu}$ at 68\% CL (shaded) and 
95\% CL (dotted lines) for ``high-D" and for ``low-D" as a function 
of the systematic offset ($\Delta$Y$_{\rm sys}$) in the value of 
Y$_{\rm P}$ derived from H II region data.  (From Hata et al. 1996b.)} 
\label{fig-4}
\end{figure}
For $\Delta$Y$_{\rm sys}$ $>$ 0.01, N$_{\nu}$ $>$ 3.0 at ``95\% CL" 
and SBBN is no longer consistent with the data; a new D/$^4$He 
``tension" appears.  In contrast, in the absence of any systematic 
offset in Y$_{\rm P}$, the $\eta$ range identified by ``low-D" 
is not consistent with SBBN (see Fig. 4).  However, as 
$\Delta$Y$_{\rm sys}$ is increased beyond 0.01, a good 
fit to SBBN may be achieved.  Thus, with increasing 
$\Delta$Y$_{\rm sys}$ consistency for SBBN shifts from 
``high-D" to ``low-D".

\subsection{Primordial D and the Baryon Density}

The conundrum posed by the contradictory results for primordial 
deuterium derived from the embryonic studies of QSO absorption-line 
systems should be resolved as more observational data are accumulated.  
Until then we may search for clues elsewhere.  Consistency for SBBN 
requires not only that the predicted primordial abundances agree with 
those inferred from observations, but also that the corresponding 
range of nucleon density identified by BBN agree with 
other astrophysical/cosmological data.  In Figure 5 (Hata 
et al. 1996b) are shown the ranges in $\Omega_{\rm B}$ as a 
function of the Hubble parameter $H_{0}$ corresponding to 
``low-D" and to ``high-D". 
\begin{figure}
\psfig{file=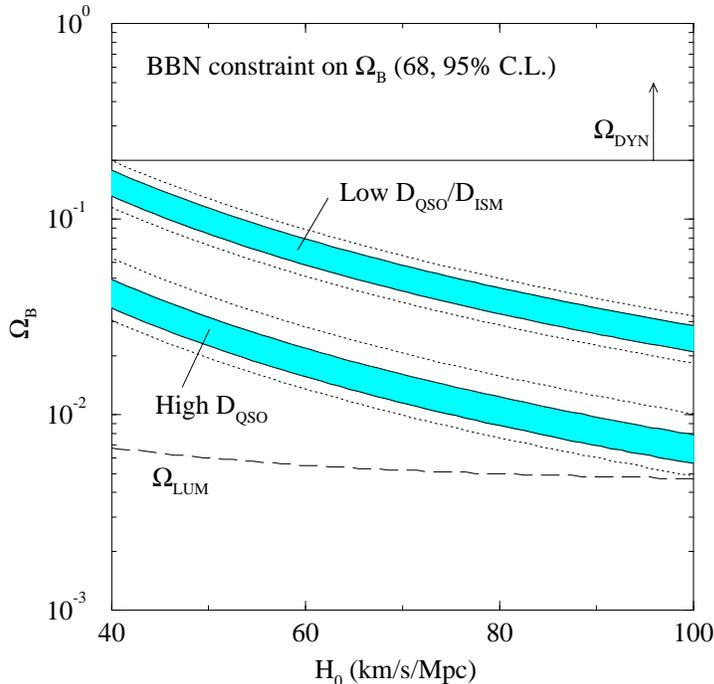,height=3.7in,width=4.0in}
\caption{The baryon density parameter ($\Omega_{\rm B}$) as a 
function of the Hubble parameter ($H_{0}$) for the 68\% CL 
(shaded) and the 95\% CL (dotted) ranges corresponding to 
``high-D" and ``low-D".  The upper bound on the density of 
``luminous" matter is shown by the dashed line and the solid 
line represents the lower bound to the total mass density.  
See the text for details and references. (From Hata et al. 1996b.)} 
\label{fig-5}
\end{figure}
Also shown is a lower bound to the baryon density inferred from 
observations of the amount and distribution of ``luminous" matter 
($\Omega_{\rm LUM}$) in the Universe (Salucci \& Persic 1996)
along with a lower bound to the total mass-energy density 
($\Omega_{\rm DYN}$) derived from the dynamics of groups 
and clusters of galaxies and from large-scale flows 
(Ostriker \& Steinhardt 1995).  Both ``low-D" and 
``high-D" $\eta$ ranges are consistent with these bounds 
and both leave room for non-baryonic dark matter 
($\Omega_{\rm B}$ $<$ $\Omega_{\rm DYN}$) and for 
dark baryons ($\Omega_{\rm B}$ $>$ $\Omega_{\rm LUM}$).  
However, the low values of $\Omega_{\rm B}$ implied by 
the ``high-D" data do provide a hint of a problem.
Is such a low baryon density really consistent with all 
the observational data?

\subsection{X-Ray Clusters and the Baryon Density}
\begin{figure}
\psfig{file=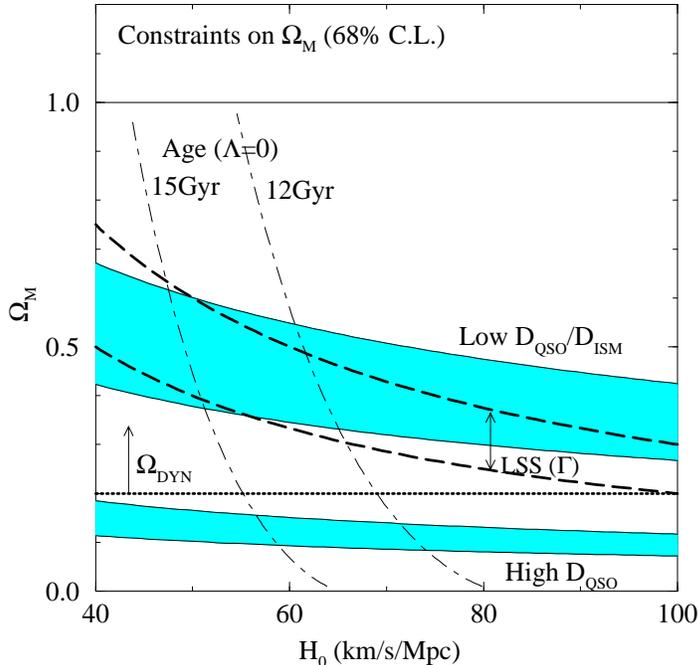,height=3.7in,width=4.0in}
\caption{Upper bounds on the total mass density parameter 
($\Omega_{\rm M}$) as a function of the Hubble parameter ($H_{0}$) 
for ``high-D" and ``low-D" (shaded bands at 68\% CL) along with 
constraints from the ``shape parameter" ($\Gamma = \Omega_{\rm M} h$) 
and from dynamics ($\Omega_{\rm DYN}$).  This figure is for a zero 
cosmological constant ($\Lambda$ = 0).  Also shown are the loci of 
$\Omega_{\rm M}$ vs. $H_{0}$ for two choices of the present 
age of the Universe.  (From Hata et al. 1996b.)} 
\label{fig-6}
\end{figure}

Rich clusters of galaxies are expected to provide a ``fair" sample 
of the universal baryon fraction ($f_{\rm B}$ $\equiv$ 
$\Omega_{\rm B}$/$\Omega_{\rm M}$); see, e.g., White \& Frenk (1991); 
White et al. (1993); White \& Fabian (1995); Evrard et al. (1995).  If, 
indeed, $f_{\rm B}$ is equal to the ratio of the cluster baryonic mass 
($M_{\rm B}$) to the total gravitating mass ($M_{\rm TOT}$), x-ray 
observations of rich clusters may be used to infer $\Omega_{\rm M}$ from 
$f_{\rm B}$ and $\Omega_{\rm B}$ (or $\eta$).  The x-ray flux provides 
information on the amount of hot gas (baryons) $M_{\rm HG}$ while the 
spectrum (x-ray temperature) is a probe of the depth of the potential well 
(under the assumption of hydrostatic equilibrium) and, hence, of 
$M_{\rm TOT}$.  Since some cluster baryons are in the luminous 
galaxies and there may be dark baryons in cluster MACHOS (Gould 
1995), $f_{\rm B}$ $>$ $f_{\rm HG}$ = $M_{\rm HG}$/$M_{\rm TOT}$, 
resulting in the bound $\Omega_{\rm M}$ $<$ $\Omega_{\rm B}$/$f_{\rm HG}$.  
From the analysis of Evrard et al. (1995) and from Evrard (1995, 
Private Communication), $f_{\rm HG}$ = (0.07 $\pm$ 0.01)$h^{-3/2}$, 
so that

\begin{equation}
\Omega_{\rm M} h^{1/2} < (0.052 \pm 0.008)\eta_{10}.
\end{equation}
In Figure 6 (Hata et al. 1996b) this upper bound on $\Omega_{\rm M}$ 
is plotted versus $H_{0}$ for the $\eta$ ranges singled out by the QSO 
``low-D" and ``high-D" determinations.  Also shown in Figure 6 is the 
``dynamical" lower bound from Figure 5 along with the large scale 
structure constraint from the ``shape parameter" 
$\Gamma = \Omega_{\rm M} h = 0.25 \pm 0.05$ (Peacock \& Dodds 1994).  
For a cosmology without a cosmological constant ($\Lambda$ = 0) the 
loci of $\Omega_{\rm M}$ versus $H_{0}$ for two choices of the present 
age of the Universe are shown as well.  The conclusion to be 
drawn from Figure 6 is that these cosmological constraints 
favor the ``high-$\eta$" range consistent with ``low-D" and 
avoid the range corresponding to ``high-D".
A similar result obtains for ``flat" cosmologies with a 
cosmological constant ($\Omega_{\rm M} + \Omega_{\Lambda} = 1$) 
or for mixed, ``hot-plus-cold" dark matter ($\Omega_{\rm B} + 
\Omega_{\rm CDM} + \Omega_{\rm HDM} = 1$) models (Hata et al. 
1996b).  The x-ray cluster, large-scale structure and dynamical 
data prefer ``high-$\eta$, low-D". 
 
\section{Summary}

Over the years primordial nucleosynthesis has provided spectacular 
support for the standard hot big bang cosmology.  For a nucleon 
abundance consistent with present density determinations, the 
SBBN-predicted abundances of D, $^3$He, $^4$He and $^7$Li, which 
span some nine orders of magnitude, agree qualitatively with the 
primordial abundances inferred from observational data.  This 
success has inspired careful quantitative analyses which serve 
as a probe of consistency and of new physics beyond the standard 
models of particle physics and cosmology.
These recent investigations have uncovered some potential problems.  
In particular, the primordial abundances of deuterium and of 
helium-4 inferred from observations of the ISM and the solar 
system (for D) and of extragalactic H II regions (for $^4$He) 
appear to be inconsistent with the predictions of SBBN.  Either 
the derived values of X$_{2\rm P}$ or of Y$_{\rm P}$ (or both) 
are wrong or SBBN should be modified.  The sense of the discrepancy 
is that the primordial abundances of D and/or $^4$He should have 
been underestimated, or the early universe should have expanded
more slowly.  
The latter possibility could be achieved if the tau neutrino
were 
very massive (10 -- 20 MeV) and unstable (0.01 -- 1 sec); this 
option can be (is being) tested at current accelerators.  The 
large body of accurate observations of $^4$He in low metallicity 
H II regions suggests it is unlikely that Y$_{\rm P}$ has been 
underestimated due to a statistical fluctuation and the small 
dispersion of the data suggests that many of the potential 
sources of systematic error may be negligible.  Nevertheless, 
it is difficult to entirely exclude a systematic offset of 
unknown origin.  In contrast, until recently the deuterium 
observations have been restricted to ``here and now" requiring 
the adoption of some assumptions/constraints on chemical 
evolution to derive the primordial abundance.  Thus, deuterium 
emerges as a possible weak link in the chain of tests of 
the consistency of SBBN.  In the best of all worlds D might 
be observed ``there and then" avoiding the uncertain chemical
evolution corrections.  Great hopes have been raised by 
claimed observations of deuterium in a few, high redshift, 
low metallicity QSO absorption line systems.  Unfortunately, 
this field is in its infancy and the abundances derived from 
current data appear inconsistent.  These data have led to a 
deuterium dichotomy, identifying not one but two, mutually 
exclusive (if the claimed statistical uncertainties are correct), 
primordial abundances for D.  ``High-D", corresponding to a 
low range for $\eta$, is entirely consistent with the 
predictions of SBBN and the inferred primordial abundances 
of the other light nuclides.  Nonetheless, this choice requires 
that the Galaxy has been very (too?) efficient in cycling gas 
through stars to reduce the primordial abundance of D to its 
observed value ``here and now".  Also, the low baryon density 
implied by ``high-D" and SBBN may be problematic.  In contrast, 
``low-D" exacerbates the already threatening crisis for SBBN, 
favoring a high range for $\eta$ corresponding to SBBN yields 
of $^4$He and $^7$Li which appear to exceed the primordial 
abundances derived from the observational data.  This higher 
range for $\eta$ may also be preferred by other 
observational data of x-ray clusters and large-scale 
structure.  As the approach to primordial deuterium via 
high-z, low-Z QSO absorbers grows from infancy to maturity, 
it is hoped that these contradictions will be resolved.  
Whether the answer will be ``high-D", ``low-D", or 
something else is anyone's guess.

\acknowledgments

I am pleased to thank Professor Sueli Viegas and her colleagues 
for their efficient organization of this stimulating and successful 
workshop and for the gracious hospitality they have provided.  
The work described here is the outgrowth of many collaborations 
and I wish to acknowledge with gratitude Sid Bludman, Dave 
Dearborn, Jim Felten, Naoya Hata, Paul Langacker, Keith Olive, 
Bob Scherrer, Dave Thomas, Monica Tosi and Terry Walker.  On 
several of the topics covered here I have profited from 
conversations with Gus Evrard, Andy Fabian, 
Evan Skillman, Sueli Viegas and Simon White.  This work is 
supported at Ohio State by the DOE (DE-AC02-76ER-01545).  This 
paper was written while I was a visitor at the Instituto 
Astronomico e Geofisico of the University of Sao Paulo and 
I am pleased to thank them for hospitality.

\end{document}